\documentclass[twocolumn,twocolumn]{IEEEtran}
\usepackage[T1]{fontenc}
\usepackage[latin9]{inputenc}
\usepackage[active]{srcltx}
\usepackage{color}
\usepackage{array}
\usepackage{amstext}
\usepackage{graphicx}
\usepackage[unicode=true,
 bookmarks=true,bookmarksnumbered=true,bookmarksopen=true,bookmarksopenlevel=1,
 breaklinks=false,pdfborder={0 0 0},pdfborderstyle={},backref=false,colorlinks=false]
 {hyperref}
\hypersetup{pdftitle={Your Title},
 pdfauthor={Your Name},
 pdfpagelayout=OneColumn, pdfnewwindow=true, pdfstartview=XYZ, plainpages=false}
\usepackage{breakurl}

\makeatletter

\providecommand{\tabularnewline}{\\}

\usepackage[caption=false,font=footnotesize]{subfig}
\usepackage{algorithm}
\usepackage{algorithmic}

\usepackage{multirow} 
\usepackage{amsmath} 
\usepackage{xcolor}

\allowdisplaybreaks[4]

\ifCLASSOPTIONcompsoc
\usepackage[caption=false,font=normalsize,labelfont=sf,textfont=sf]{subfig}
\else
\usepackage[caption=false,font=footnotesize]{subfig}
\fi

\usepackage{cite}
\usepackage{bm}
\usepackage{algorithmic}
\usepackage{algorithm}
\usepackage{graphicx}
\IEEEoverridecommandlockouts

\usepackage{lettrine}

\@ifundefined{showcaptionsetup}{}{%
 \PassOptionsToPackage{caption=false}{subfig}}
\usepackage{subfig}
\makeatother

\begin{document}

\title{\textcolor{black}{Column Generation for Optimization Problems in
Communication Networks}}

\author{Ziye Jia, \IEEEmembership{Member,~IEEE}, Qihui Wu, \IEEEmembership{Senior Member,~IEEE},
Chao Dong, \IEEEmembership{Member,~IEEE}, \\Chau Yuen, \IEEEmembership{Fellow,~IEEE},
and Zhu Han, \IEEEmembership{Fellow,~IEEE}\thanks{Ziye Jia, Qhui Wu and Chao Dong are with the College of Electronic
and Information Engineering, Nanjing University of Aeronautics and
Astronautics, Nanjing 211106, China, (e-mail: jiaziye@nuaa.edu.cn,
wuqihui@nuaa.edu.cn, dch@nuaa.edu.cn).

Chau Yuen is with the Engineering Product Development Pillar, Singapore
University of Technology and Design, Singapore (e-mail: yuenchau@sutd.edu.sg).

Zhu Han is with the University of Houston, TX 77004, USA (e-mail:
zhan2@uh.edu), and also with the Department of Computer Science and
Engineering, Kyung Hee University, Seoul, 446-701, South Korea.

\textit{\textcolor{black}{}}}}
\maketitle
\begin{abstract}
Numerous communication networks are emerging to serve the various
demands and improve the quality of service. Heterogeneous users have
different requirements on quality metrics such as delay and service
efficiency. Besides, the networks are equipped with different types
and amounts of resources, and how to efficiently optimize the usage
of such limited resources to serve more users is the key issue for
communication networks.\textcolor{black}{{} One powerful mathematical
optimization mechanism to solve the above issue is column generation
(CG), which can deal with the optimization problems with complicating
constraints and block angular structures. In this paper, we first
review the preliminaries of CG}\textcolor{blue}{.}\textcolor{black}{{}
Further, the branch-and-price (BP) algorithm is elaborated, which
is designed by embedding CG into the branch-and-bound scheme to efficiently
obtain the optimal solution. The applications of CG and BP in various
communication networks are then provided, such as space-air-ground
networks and device-to-device networks. }In short, our goal is to
help readers \textcolor{black}{refine} the applications of the CG
optimization tool in terms of problem formulation and solution. We
also discuss the possible challenges and prospective directions when
applying CG in the communication networks.
\end{abstract}

\section{Introduction}

With the increment of various requirements in the 5G and beyond techniques,
heterogenous communication networks, such as the space-air-ground
networks, multi-access edge computing (MEC) networks, device-to-device
(D2D) networks, Internet of vehicles, satellite networks, and body
area networks, are constantly emerging to support different applications.
A significant challenge in these communication networks is the optimization
decision-making, to satisfy a large number of users' requests with
various quality of service (QoS) metrics, and at the same time to
maximize network performance and resource efficiency. Specific categories
of the optimization problem in these networks include routing decision,
virtual network function deployment, resource competition, etc. The
optimization decision is based on the characteristics of communication
networks, as well as QoS demands of various users. Consequently, it
is significant to design efficient algorithms for the optimization
decision problem in communication networks.

\begin{figure*}[t]
\centering

\includegraphics[scale=0.52]{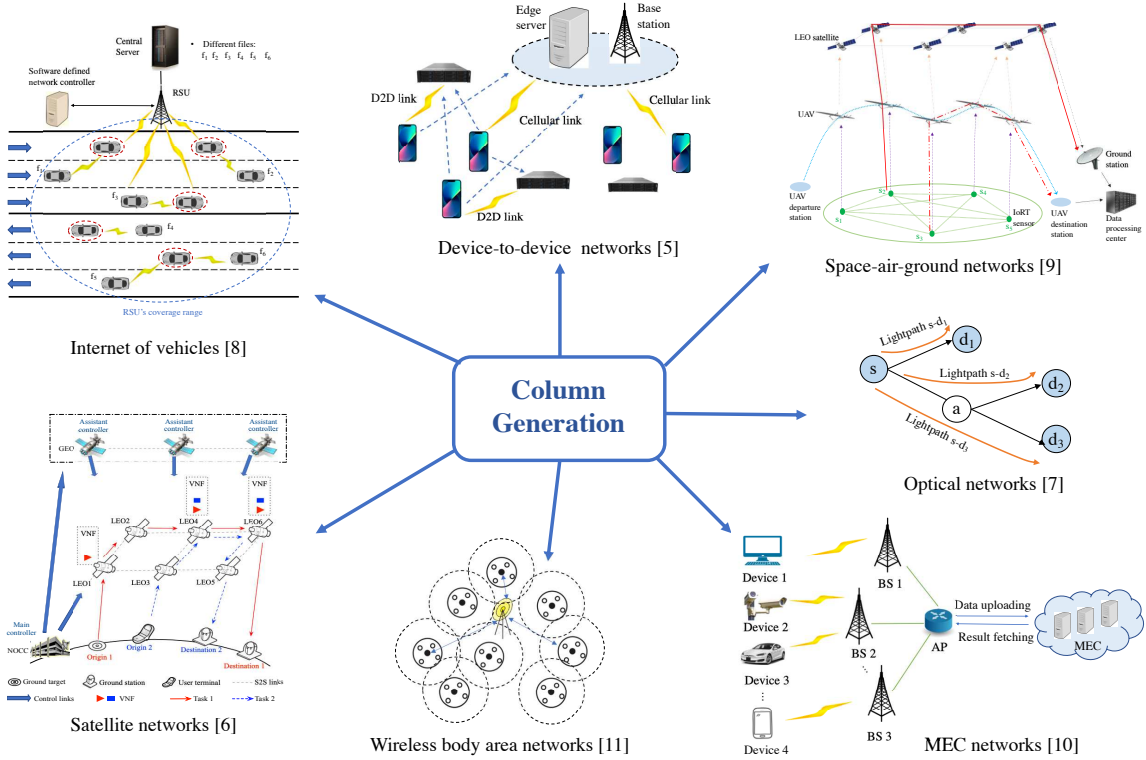}\textcolor{black}{\caption{\textbf{\textcolor{blue}{\label{fig:CG-case}}}\textbf{\textcolor{black}{{}
}}\textcolor{black}{Use cases of column generation in communication
networks.}}
}
\end{figure*}

\textcolor{black}{Regardless of the network types, mathematical programming
has extensive applications in the optimization decision problems.
The decision problem has a certain optimization objectives (e.g.,
energy utilization), a couple of decision variables (e.g., integer
or linear), and is restricted by multiple constraints (e.g., resource
capacity constraints). With the increment of network scale, both the
number of decision variables and the number of constraints will increase,
which results in unacceptable computation complexity in practice for
the optimal decision. However, most problems have the special structure
within the decision variables or constraints, and if the complicating
variables or constraints are separated, the decomposition scheme will
be available. As for the optimization problem with complicating constraints
and block angular structures, column generation (CG) is applicable
for acquiring a suboptimal solution with high efficiency.}

\textcolor{black}{As one of the effective decomposition methods, CG
is based on the Minkowski's theory and Dantzig-Wolfe decomposition
\cite{155-A-primer-in-CG}, via decomposing the original intractable
problem into a master problem (MP) and a couple of independent pricing
problems (PPs).} Compared with the original problem, both the MP and
PP have smaller scale variables and constraints, and can be handled
with a lower computational burden. In particular, CG leverages the
block angular structure of the original problem when ignoring the
complicating constraints, and the complicating constraints will be
handled in MP. CG has been widely used in various communication networks,
as the use cases shown in Fig. \ref{fig:CG-case}.\textcolor{black}{{}
In this work, we provide the preliminaries and recent applications
of CG, as well as the analyses and tricks to deal with the MP and
PP in different communication networks. Besides, since CG may lost
the optimality during the iteration between MP and PP, we also introduce
the branch-and-price (BP) scheme \cite{200-Branch-and-price}, which
works by embedding CG into the classical branch-and-bound framework
to efficiently guarantee the optimal solution. Finally, a variety
of challenges and possible schemes are analyzed for future research
directions. }

\begin{figure*}[tbh]
\centering

\includegraphics[scale=0.39]{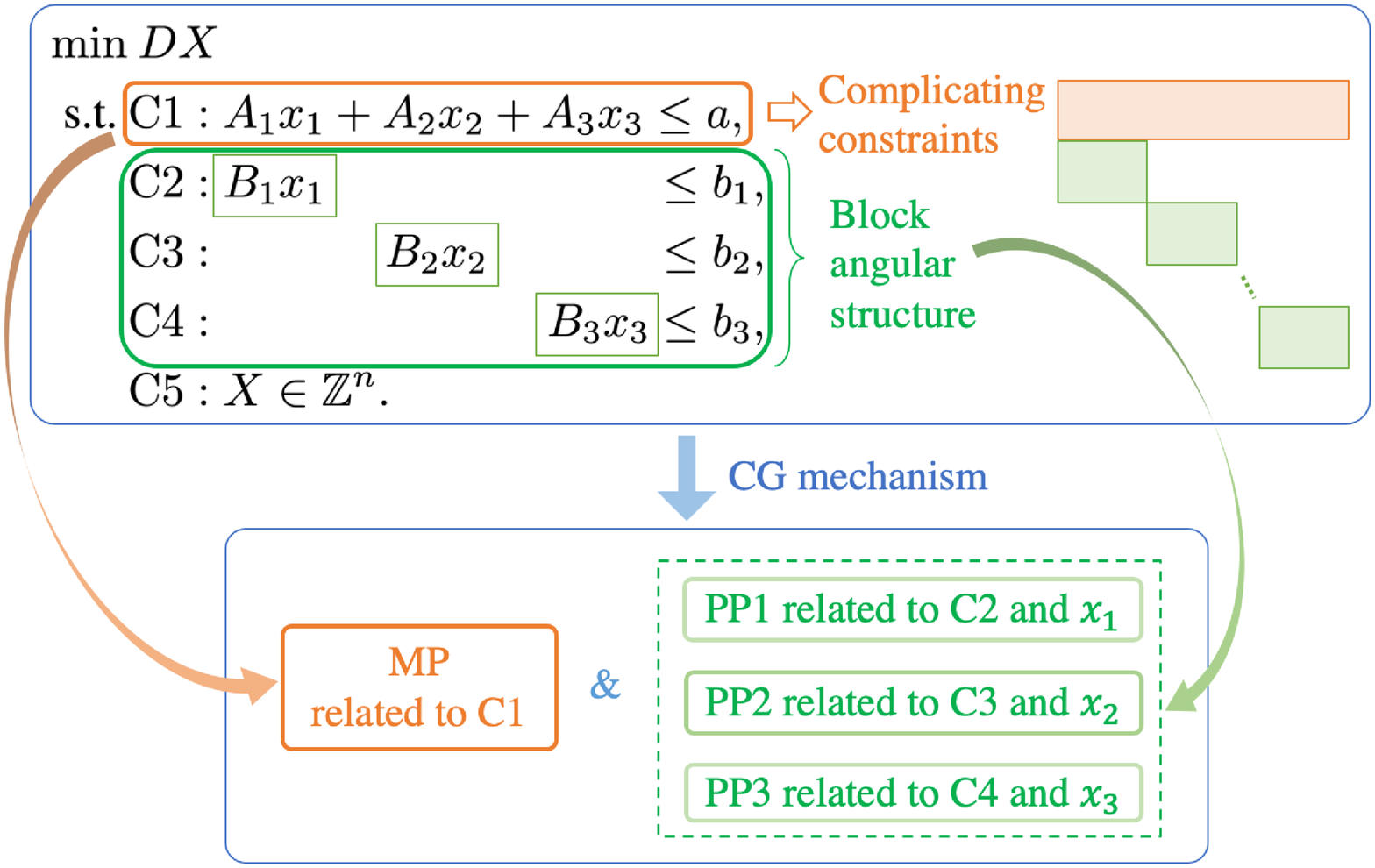}\caption{\textcolor{blue}{\label{fig:block_angular}}\textcolor{black}{A view
of the optimization problem with block angular structure and the decomposition
via CG.}\textcolor{blue}{{} }}
\end{figure*}
\begin{figure*}[tbh]
\centering

\subfloat[\label{fig:Flowchart-CG}Flowchart of CG.]{\centering

\includegraphics[scale=0.6]{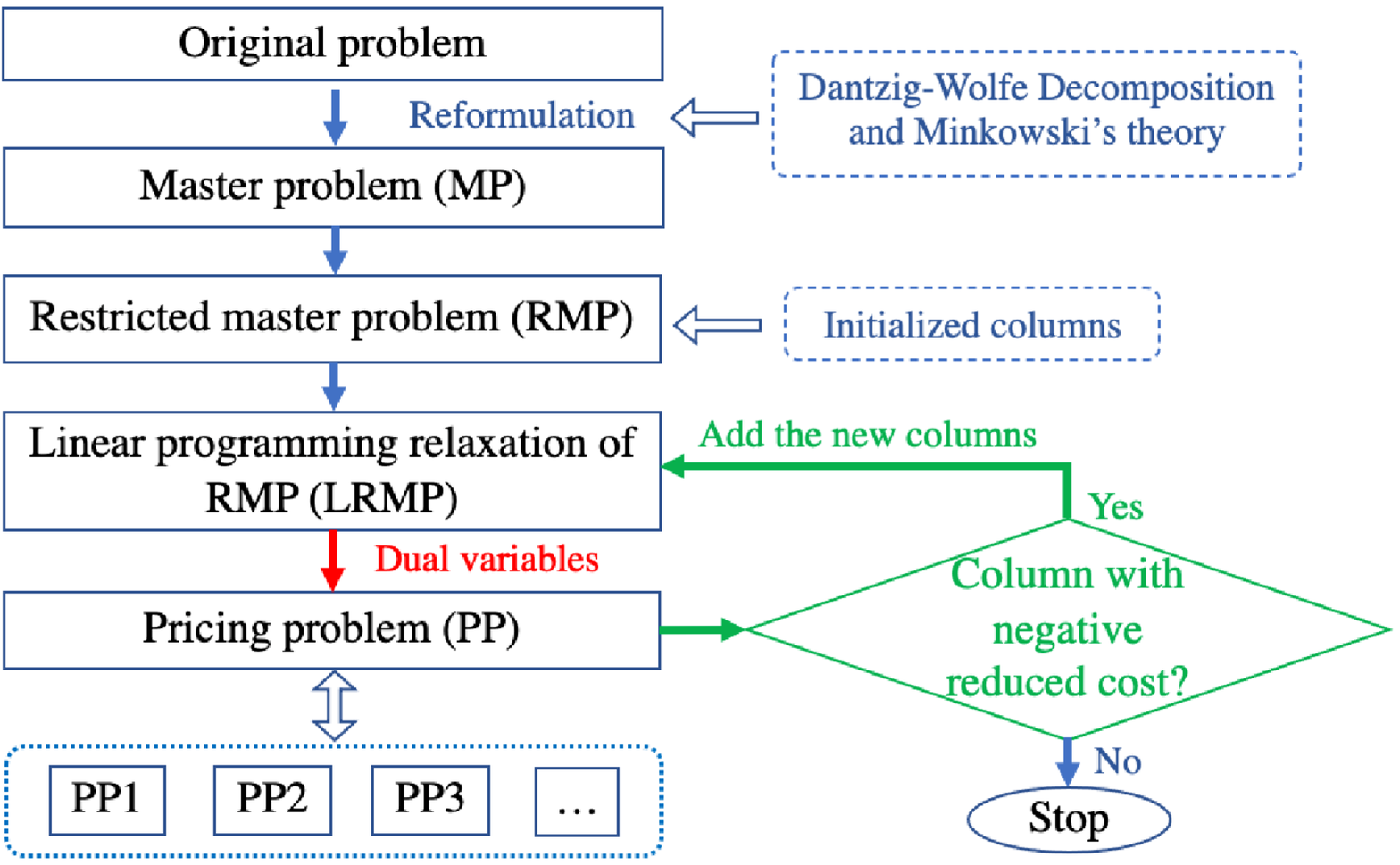}\textcolor{black}{}}

\subfloat[\label{fig:Iteration-MP-PP} Iteration between MP and PP.]{\centering

\includegraphics[scale=0.48]{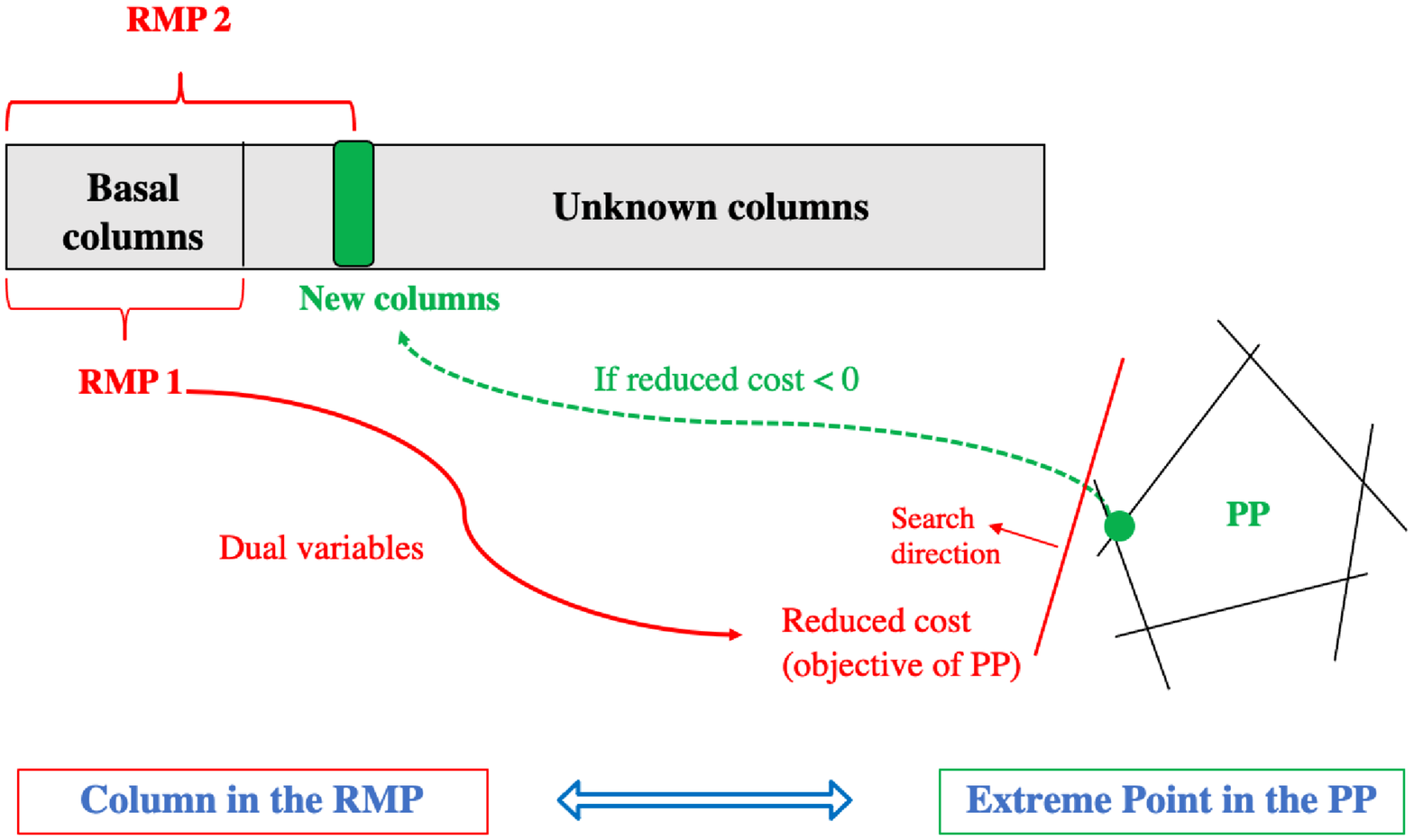}\textcolor{black}{{} }}\caption{\textcolor{blue}{\label{fig:CG}}\textcolor{black}{Illustration of
CG mechanism.}}
\end{figure*}

\section{Basics of Column Generation}

\subsection{Preliminary of Column Generation}

\textcolor{black}{Primarily, CG is applicable for the optimization
problem in the form of linear programming and integer programming
with complicating constraints \cite{778-CG}. The premise of CG is
that the remaining constraints of the original problem are in the
block angular structure if the complicating constraints are dropped.}
The remaining problem can be tackled in parallel, while the complicating
constraints can be handled\textcolor{blue}{{} }\textcolor{black}{in
MP} \cite{675-Selected-topics-in-CG}. In detail, the basic idea behind
CG is decomposing the original problem into two smaller scale problems:
a MP and a serious of PPs, and the iteration between MP and PPs will
finally obtain the solution for the original problem. 

\textcolor{black}{Moreover, to help readers without background on
this area to better understand the technique, a simple and general
optimization problem with complicating constraints and block angular
structure is illustrated in Fig. \ref{fig:block_angular}, to clarify
the problem properties such as the block angular structure, MP and
PP. Specifically, from the optimization problem in Fig. \ref{fig:block_angular},
it is noted that C1 is a complicating constraint since it is related
to all variables $x_{1}$, $x_{2}$, and $x_{3}$, while constraints
C3-C4 are respectively related to different variables, and such a
problem structure is known as the block angular structure \cite{200-Branch-and-price}.
An optimization problem with constraints in the form of block angular
can be handled by the CG mechanism with efficiency, since the complicating
constraints can be removed and only handled in the MP, while the residual
constraints are independent with different variables, which forms
mutually independent PPs. For instance, in Fig. \ref{fig:block_angular},
PP1 is only related to constraint C2 and $x_{1}$, and PP2 is only
related to constraint C3 and $x_{2}$. Hence, these PPs can be solved
concurrently, which further improves the solution efficiency.}

\begin{figure*}[tbh]
\centering

\includegraphics[scale=0.6]{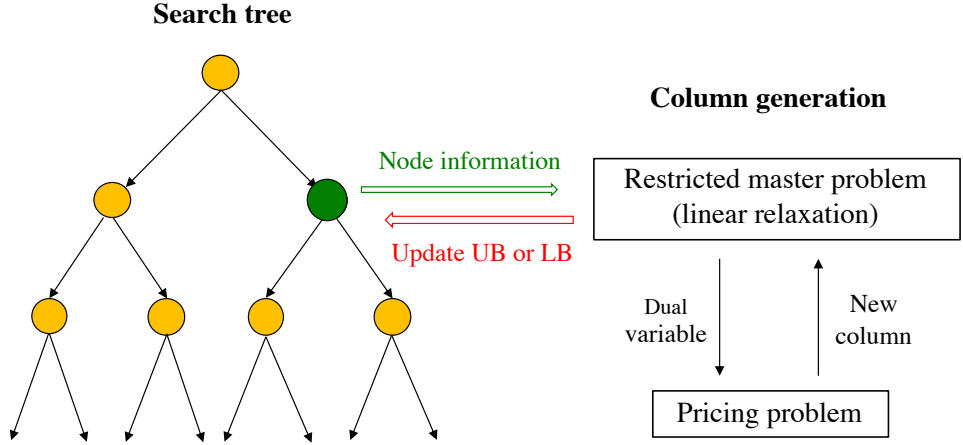}\textcolor{black}{\caption{\label{fig:BP-view}A view of the branch-and-price mechanism. }
}
\end{figure*}
\textcolor{black}{Fig. \ref{fig:CG} provides the detailed CG mechanism,
including the flowchart of CG in Fig. \ref{fig:Flowchart-CG}, and
a vivid illustration of the iteration between the RMP and PP in Fig.
\ref{fig:Iteration-MP-PP}. Firstly, in Fig. \ref{fig:Flowchart-CG},
the original problem is reformulated by transforming the decision
variables according to the Minkowski's theory, i.e., a point in the
convex hull of the original problem can be expressed as the linear
combination of the extreme points of the convex hull \cite{155-A-primer-in-CG}.
Consequently, the extreme points become the decision variables of
PP, and the linear combination coefficients turn into the decision
variables of MP. Such a separation is known as the Dantzig-Wolfe decomposition.
Since the enumeration of all columns for MP is impossible, just a
fraction of active columns is considered, which is called the restricted
master problem (RMP). To activate the iteration between RMP and PP,
RMP should be initialized to obtain the basic columns, and such initialization
can be implemented via heuristic algorithms according to the property
of different networks. Besides, the initialization just needs to obtain
a set of feasible solutions for RMP, and good initialization for RMP
will reduce the iteration times between RMP and PP. }

If RMP is a linear programming problem, its dual variables are directly
calculated. Otherwise, if there exist integer variables in RMP, the
linear programming relaxation of RMP (LRMP) is calculated, and the
dual variables are obtained. For clarity, in Fig. \ref{fig:Flowchart-CG},
the general case that RMP is not linear programming is considered.
Then, the dual variables of LRMP are provided to PP to update the
optimization object of PP, which is also named as \textit{\textcolor{black}{reduced
cost}}, and PP can be solved directly by the optimization tools, such
as CVX and MOSEK, to obtain the solutions and objective value (\textit{reduced
cost}). \textcolor{black}{The solutions of PP are known} as new columns,
and the function of PP is used for generating new columns. After that,
the value of reduced cost is verified to decide whether the new columns
will be added to LRMP. Specifically, if the reduced cost is negative,
the columns are added to RMP. Otherwise, the iteration between LRMP
and PP is terminated, since in this case, the solution of LRMP cannot
be further improved by adding new columns from PP. 

At this point, the current LRMP is solved with the existing columns
to obtain the final solutions. Note that if RMP is equivalent to LRMP,
i.e., RMP is a linear programming problem, the final solution of LRMP
is the optimal solution of the original problem, since in this case,
the original problem is equivalent to RMP. Besides, if RMP has integer
decision variables and the final solutions of LRMP are integer values,
the solutions are also the optimal solutions of RMP as well as the
original problem. Otherwise, if the final solutions of LRMP are not
integer, approximated methods should be employed to deal with the
fractional solutions into integer. As such, the optimality may be
lost and only the sub-optimal solutions are obtained\textcolor{black}{.
If the optimal solution need to be guaranteed, the BP algorithm should
be applied here, which will be elaborated later in Section \ref{sec:Branch-and-Price}.}
\textcolor{black}{In addition, RMP has a better structure than the
original problem, since compared with directly relaxing the original
problem, the linear programming relaxation of RMP will yield a tighter
bound \cite{1043-fog-D2D-BP}.}

Moreover, to clearly figure out the iteration between RMP and PP,
Fig. \ref{fig:Iteration-MP-PP} provides a vivid illustration. We
can observe that the basic columns are initialized for the first RMP
(RMP 1), and other solution set is still undiscovered as unknown columns.
Then, the dual variables are calculated to help update the objective
(reduced cost) of PP, and meanwhile, the search direction of PP is
also decided. After PP being solved, the reduced cost is verified.
If the reduced cost is less than 0, the new columns generated by PP
are added to RMP, and the column number of RMP is widen. Then, the
second RMP (RMP 2) is calculated and a new iteration between RMP and
PP \textcolor{black}{starts}. Besides, it is observed that the solutions
of PP, i.e., the extreme points in PP, are corresponding to the columns
in RMP, which explains why PP can help generate new columns for RMP.
In addition,\textcolor{blue}{{} }\textcolor{black}{it is not necessary
to obtain the optimal solution for PP in each iteration, and only
the columns with negative reduced cost will be applicable. Hence,
an approximation algorithm for PP with efficiency will be significant. }

\subsection{Branch-and-Price for \textcolor{black}{Optimality}\label{sec:Branch-and-Price}}

If there exists integer decision variables in the original problem,
the final results from CG may lost the optimali\textcolor{black}{ty.
To guarantee the optimality, BP is introduced by embedding CG into
the branch-and-bound framework. Compared with the exhaustive search,
BP will help obtain the optimal solution more efficiently, and there
exist no optimality gap between the solution of BP and the optimal
solution. As such, BP can serve as the benchmark for other sub-optimal
algorithms. To illustrate the process of BP, Fig. \ref{fig:BP-view}
provides a view for BP mechanism. In detail, since the key point of
BP is that CG is inserted in the branch-and-bound scheme, a search
tree should be generated. By the way, the branch-and-bound framework
works by building a search tree, and exploring the branches of the
search tree to update the lower bound (LB) or upper bound (UB) until
convergence. Note that the search node in the search tree is corresponding
to the decision variable in the original problem, i.e., BP works branch
at the search node with an original variable. The branch strategy
on the decision variable of RMP is useless, since if any column is
cut when branch, the column may be regenerated by PP as a new column,
resulting in the endless loop \cite{200-Branch-and-price}. Then,
at each search node, CG is implemented, and the LB or UB is updated
after executing CG. For example, in \cite{JZY-TWC}, if the solution
of LRMP is integer and larger than the UB, the current node and its
corresponding subtree are pruned. If the solution of LRMP is integer
and less than the incumbent UB, UB is updated as the solution of LRMP,
and the current search node is fathomed. If the solution of LRMP is
fractional, it is served as the LB. The renewed UB and LB will help
to cut the search tree, and obtain the final optimal solution when
UB equals to LB or the search tree is empty. Compared with the traditional
branch-and-bound scheme, due to CG, the solution of each node will
be efficient, and the search tree will be cut faster.}

\textcolor{black}{Besides, the initial values of LB or UB can be obtained
by a heuristic mechanism according to different network characteristics.
For instance, in the satellite networks of \cite{JZY-TWC} with a
minimized optimization objective, the initial value of UB is obtained
by a heuristic algorithm for RMP, since the UB is a feasible solution
for the original problem, and LB is initialized as 0. Interested readers
are recommended to the specific procedures of BP algorithm in \cite{1043-fog-D2D-BP}
and \cite{JZY-TWC}.}

\textcolor{black}{In regard to the time complexity, according to \cite{JZY-TWC},
the complexity of brute-force searching for the original integer programming
problem is $\mathcal{O}(M_{1}\cdot N_{1}\cdot2^{M_{1}})$, in which
$M_{1}$ and $N_{1}$ are the number of variables and constraints,
respectively. When CG is applied, LRMP can be directly solved by the
optimization tools, and the time-consuming procedure is the solution
for PP, which is a small scale integer programming problem. The time
complexity of PP is $\mathcal{O}(M_{2}\cdot N_{2}\cdot2^{M_{2}})$,
in which $M_{2}$ is the number of variables and $N_{2}$ is the number
of constraints of PP. Although it is still with the exponential complexity,
due to that $M_{2}\ll M_{1}$ and $N_{2}\ll N_{1}$, the time complexity
is significantly reduced. As for the time complexity of BP, in the
worst case, it is $\mathcal{O}(M_{1}\cdot N_{1}\cdot2^{M_{1}})$.
However, when CG is applied, the solution of each search node of the
search tree will be faster, and the LB and UB will converge quickly
due to the good initialization in CG. }

\section{\textcolor{black}{Applications in Communication Networks}}

\textcolor{black}{CG has been employed in various communication network
optimization design with the complicating constraints and block angular
structures. Moreover, to acquire the optimal solution, BP is applied
as an efficient scheme in these communication networks. A couple of
use cases of CG and BP are listed in Table \ref{tab:CG-cases}. }

\subsubsection{CG for Optical Networks}

In \cite{254-CG-opticalNet}, CG is applied for the multicast provision
in the mixed-line-rate optical networks, with the objective of minimizing
the total cost of transponder, wavelength channel, and employed wavelength
number. Three types of binary integer variables and a continuous variable
are employed in the system model, and the light path based multicast
provision problem is formulated in the form of integer linear programming.
To efficiently tackle the NP-hard problem, CG is employed to decompose
the original problem into the PP of potential light path generation
problem and the RMP to select the final solution light path with the
minimum total cos\textcolor{black}{t. Besides, for acceleration, the
elementary shortest path with resource constraints algorithm is applied
in both initialization of RMP and PP, and the conflict graph based
method for wavelength allocation is designed to efficiently obtain
the final solution.}

\subsubsection{CG for Internet of Vehicles}

\cite{1037-vehicularN-CG} investigates the joint frequency scheduling
and power control for the Internet of vehicle networks, to maximize
the total number of tuple links in a given scheduling time period.
The specific problem is in the form of mixed-integer nonlinear programming
with a binary variable indicating if a tuple link is applied for transmission
and a continuous variable denoting the transmitting power. CG is employed
with the RMP of a transmission power scheduling problem and PP of
power control for tuple-links of each pattern\textcolor{black}{. To
accelerate the solution of CG, a greedy power allocation method is
designed for PP.} 
\begin{table*}[p]
\centering

\caption{\textcolor{black}{\label{tab:CG-cases}Applications of column generation
and branch-and-price in various networks.}}

\begin{tabular}{|>{\raggedright}m{1.1cm}|>{\raggedright}m{1.8cm}|>{\raggedright}m{1.7cm}|>{\raggedright}m{3.8cm}|>{\raggedright}m{1.6cm}|>{\raggedright}m{1.4cm}|>{\raggedright}m{2.7cm}|}
\hline 
Scenario & Problem description & Objective & Decision variables & RMP & PP & \textcolor{black}{Improvements}\tabularnewline
\hline 
\hline 
\multicolumn{7}{|c|}{\textbf{\textcolor{black}{Column generation}}}\tabularnewline
\hline 
Optical network \cite{254-CG-opticalNet} & Multicast provision in mixed-line-rate optical networks. & Minimize the total cost and the number of used wavelength. & $\cdot$ Binary variable denoting whether a light path uses the link.
\\

$\cdot$ Binary variable representing whether a wavelength is used.\\
$\cdot$ Binary variable indicating whether a wavelength is used by
the link.\\
$\cdot$ \textcolor{black}{Continuous variable denoting the distance
from a node to the source.} & Select the light path with the minimum total cost. & Potential light path generation. & \textcolor{black}{$\cdot$ The elementary shortest path with resource
constraints algorithm is used in both initialization of RMP and PP.}

\textcolor{black}{$\cdot$ Conflict graph based method for wavelength
allocation.}\tabularnewline
\hline 
Internet of vehicles \cite{1037-vehicularN-CG} & Joint frequency scheduling and power control for Internet of vehicle
networks. & Maximize the total number of tuple links in the scheduling time. & $\cdot$ Binary variable indicating whether a tuple-link is used for
transmission during the scheduling time.\\
$\cdot$ Continuous variable of transmitting power. & Transmission pattern scheduling problem. & Power control for tuple-links of each pattern. & \textcolor{black}{$\cdot$ A greedy power allocation method is applied
for PP.}\tabularnewline
\hline 
Space-air-ground network \cite{jzy-IoTJ2020} & Joint IoT data collection and trajectory of UAV assisted by low earth
orbit satellites.  & Minimize the total energy cost of UAV. & $\cdot$ Binary variable denoting whether an IoT device connects a
UAV.\\
$\cdot$ Binary variable denoting whether a UAV passes a virtual arc. & UAV trajectory path decision. & Potential trajectory generation for a UAV. & \textcolor{blue}{$\cdot$ }\textcolor{black}{A heuristic algorithm
designed for the initialization of RMP as a warm start.}

\textcolor{black}{$\cdot$ A shortest path based approximated trajectory
for the PP.}\tabularnewline
\hline 
MEC network \cite{yangyang-MEC-CG1} & IoT task offloading in MEC with latency requirement. & Minimize the summarized resource consumption. & $\cdot$ Binary variable denoting whether an offloading configuration
is used.\\
$\cdot$ Binary variable representing whether a virtual node is included
in the offloading configuration. & Decide the optimal computation offloading configuration. & Generate additional offloading configurations. & \textcolor{black}{N/A.}\tabularnewline
\hline 
Wireless body area network \cite{1045-bodyANet-CG} & User admission with interference in wireless body area network  & Maximize the total accessed users with time assignment. & $\cdot$ Binary variable denoting whether a user accesses the network.\\
$\cdot$ Continuous variable indicating the time assignment to feasible
candidate groups. & Select the candidate group with minimum time assignment. & \textcolor{black}{Promising candidate user group generating.} & \textcolor{black}{$\cdot$ A greedy initialization algorithm for RMP.}

\textcolor{black}{$\cdot$ Maximum weighted independent set based
approximation method for PP.}\tabularnewline
\hline 
\hline 
\multicolumn{7}{|c|}{\textbf{Branch-and-price}}\tabularnewline
\hline 
Device-to-device network \cite{1043-fog-D2D-BP} & Resource management for the device-to-device assisted fog computing. & Maximize the total management profit. & $\cdot$ Binary variable indicating if a user select a certain candidate
server and an offloading channel.\\

$\cdot$ Continuous variable of transmission power from a user allocated
on a link with a certain channel. & The optimal link subset selection. & \textcolor{black}{Candidate feasible link subset generation.} & \textcolor{black}{$\cdot$ Add additional branching constraints.\\}

\textcolor{black}{$\cdot$ A greedy algorithm of sequential multiple
resource allocation to obtain suboptimal solution.}\tabularnewline
\hline 
Satellite network \cite{JZY-TWC} & Resource allocation for software defined satellite networks. & Minimize the total communication resource consumption.  & $\cdot$ Binary variable indicating whether a virtual network function
deploys on a satellite.\\
$\cdot$ Binary variable denoting whether the data is routed by a
satellite-to-satellite link. & Service path selection for all tasks. & \textcolor{black}{Service path generation for each task.} & \textcolor{black}{$\cdot$ K-shortest path based mechanism is used
in both RMP and PP.\\ }

\textcolor{black}{$\cdot$ Beam search for branching in the search
tree.\\}

\textcolor{black}{$\cdot$ A heuristic algorithm for the last RMP.}\tabularnewline
\hline 
Data center network \cite{1044-BP-SDDCN} & Dynamic mapping of gNB to the software virtual machine pool. & Minimize cloud computing and energy cost, and maximize processed traffic
load.  & $\cdot$ Binary variable denoting whether a gNB is mapped on a virtual
machine.\\

$\cdot$ Binary variable indicating whether a virtual node is active. & Decide the optimal mapping set.  & \textcolor{black}{Generate the gNBs to VM pool mapping association.} & \textcolor{black}{$\cdot$ Generate additional cuts to RMP to shrink
the bound.}\tabularnewline
\hline 
\end{tabular}
\end{table*}

\subsubsection{CG for Space-Air-Ground Networks}

A problem of joint UAV trajectory and Internet of Things (IoT) data
collection assisted by satellite networks is studied in \cite{jzy-IoTJ2020}.
The purpose is to minimize the total energy cost of UAV with multiple
resource restrictions and flow constraints. To figure out the problem
with efficiency, CG is employed with RMP to optimize the best UAV
trajectory from the feasible sets generated by PP\textcolor{black}{.
In addition, a heuristic algorithm based on the resource constrained
shortest path is designed for the initialization of RMP as a warm
start. Also, a shortest path based approximated scheme is designed
for PP to reduce the computational complexity. }

\subsubsection{CG for MEC Networks}

In \cite{yangyang-MEC-CG1}, an IoT task offloading problem in the
MEC network with latency requirement is presented, with the target
to minimize the summarized resource consumption, and two types of
binary denoting offloading decision and offloading configuration,
which is in the form of integer programming. CG is applied for the
optimization problem by iteration between the decomposed RMP and PP.
In particular, PP is\textcolor{black}{{} used for }generating additional
offloading configurations and RMP is designed for deciding the computation
offloading configuration with the minimum weighted total cost. 

\subsubsection{CG for Wireless Body Area Networks}

\cite{1045-bodyANet-CG} focuses on the problem of user admission
with interference in wireless body area network, with the objective
of maximizing the total accessed users with time assignment. There
are two types of variables: the binary variable denoting whether a
user can access the network and the continuous variable indicating
time assignment to feasible candidate groups. The formulated problem
is in the form of mixed integer nonlinear programming. To tackle the
intractability, CG is employed. Therein, RMP is used for selecting
the candidate group with minimum time assignment, and PP generates
the promising candidate user group.\textcolor{black}{{} Besides, an
accelerated CG algorithm assisted by a greedy initialization algorithm
for RMP and a maximum weighted independent set based approximation
method for PP is further proposed for solution efficiency.}

\subsubsection{\textcolor{black}{BP for D2D Assisted Fog Computing Networks}}

In \cite{1043-fog-D2D-BP}, BP is employed to tackle the problem of\textcolor{blue}{{}
}\textcolor{black}{link scheduling, channel assignment as well as
the power control} in D2D assisted fog computing networks, which is
in the form of mixed integer nonlinear programming. In detail, at
each search node, CG is applied to efficiently obtaining the optimal
solution at the current node by RMP of selecting the optimal feasible
link subset and PP of generating the possible feasible set. Besides,
additional branching constraints are added to accelerate convergence.\textcolor{black}{{}
Further, a greedy algorithm for the original problem based on sequential
multiple resource allocation is designed to acquire suboptimal solution
with high efficiency, especially in the large fog-computing system.}

\subsubsection{BP for Satellite Networks}

A resource allocation problem for the software defined satellite networks
is presented in \cite{JZY-TWC}, with the objective to minimize the
summarized communication resource consumption in a time horizon. BP
is employed to handle the formulated problem, and in each search node,
CG is employed by RMP of selecting the optimal service path and PP
to generate the potential service paths for RMP\textcolor{black}{.
To accelerate the pruning for the branch tree, the K-shortest path
based scheme is employed to initialize RMP and solve PP. Besides,
the beam search is applied to accelerating the search tree cutting,
and a heuristic algorithm is designed for the last RMP to obtain tighter
UB.}

\begin{figure}[tbh]
\centering

\subfloat[\label{fig:timecost}Time cost.]{\centering

\includegraphics[scale=0.44]{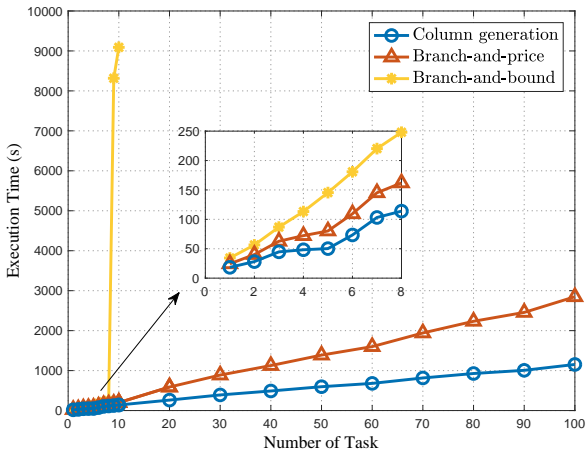}

}

\subfloat[\label{fig:results}Optimization results.]{\centering

\includegraphics[scale=0.44]{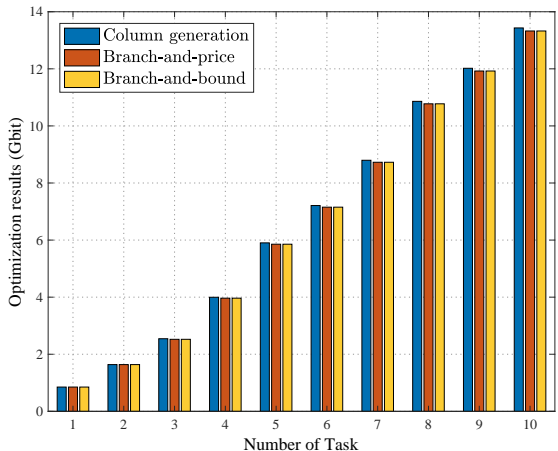}

}

\caption{\label{fig:Numerical-results}Numerical results for satellite networks
\cite{JZY-TWC}. }
\end{figure}

\textbf{Numerical example:} Due to the page limitation, we show only
the results of \cite{JZY-TWC}. Specifically, the numerical results
are conducted in the scenario of 16 low earth orbit satellite network,
constituting $\textrm{4\ensuremath{\times}4}$ Walker constellation
and each orbit is composed of 4 satellites. In particular, Fig. \ref{fig:Numerical-results}
provides the numerical results comparison of different methods in
time cost (Fig. \ref{fig:timecost}) and optimization results (Fig.
\ref{fig:results}). Except for the column generation and BP, the
optimal results via the branch-and-bound is leveraged as a benchmark.
The optimization results denote the total communication resource consumption,
which is the optimization objective of \cite{JZY-TWC}. By combining
Figs. \ref{fig:timecost} and \ref{fig:results}\textcolor{black}{,
it is observed that the result of BP can obtain the optimal solution
with low time cost, while CG can reach a near-optimal result with
lower computational complexity.}

\subsubsection{BP for Software Defined Data Center Network}

A dynamic mapping of next generation node-Bs (gNBs) to software virtual
machine pool problem is proposed in \cite{1044-BP-SDDCN}, with the
weighted objective to minimize the total cloud computing cost, processing
power and to \textcolor{black}{maximize the network traffic load}
processed by virtual machine pools. BP is employed for handling the
problem, with the PP to generate the mapping association and RMP to
decide the optimal mapping set\textcolor{black}{. Besides, to accelerate
the solution efficiency, additional cuts are added to RMP to shrink
the bound.}

\textcolor{black}{As above, CG can acquire the near optimal solution
with high efficiency in the network optimization problems discussed,
while BP guarantees the optimality. Hence, CG is applicable to large-scale
network optimization problems to obtain a satisfied solution with
requirements of low time complexity. }

\section{Challenges and Directions }

Although CG is an effective mechanism to tackle the optimization problem
in communication networks, there still exists a couple of challenges:
1) time-consuming iterations between RMP and PP as well as the time
cost of optimal solution of PP; 2) the initialization and final integer
solution decision for RMP; 3) long tail effect with the increasing
iteration times between RMP and PP; 4) CG used in the \textcolor{black}{stochastic
optimization}\textcolor{blue}{;} 5) acceleration of BP to acquire
a less elegant but more efficient solution. To tackle these challenges,
the following tips will be helpful. 
\begin{itemize}
\item As for the time-consuming iterations between RMP and PP, since the
LRMP can be directly solved by optimization tools, the computational
burden primarily comes from solving PP. Besides, a good solution for
PP will help provide high-quality columns for RMP. In particular,
the solution of PP can be acquired according to the property of PP
in detailed networks, for example, a shortest-path problem \cite{jzy-IoTJ2020}.
In addition, the optimal solution for PP will decrease the iterations,
however, to obtain the optimality is time-consuming, so there exists
a trade-off between the iteration times and the solution quality of
PP. 
\item A feasible initial solution of RMP is necessary to activate the CG
scheme \cite{200-Branch-and-price}, and a warm start will provide
high-quality solution and help reduce the total time cost, for example,
the heuristic warm initialization for RMP in \cite{254-CG-opticalNet}.
In addition, if the final solution of RMP is not integer when the
iteration between RMP and PP terminates, to address the efficiency
and solution quality of RMP, an effective approximated mechanism for
RMP is applicable. Besides, additional cuts can be generated and added
to RMP to shrink the search area.
\item The long tail effect \cite{274-crosslayerCG-LXiao} results in too
many iteration times, which can be dealt with by predefining an iteration
value, since the final solution is nearly obtained, but the convergence
speed is quite slow. 
\item In the stochastic optimization problem with uncertainty demand, CG
is available to tackle the problem with efficiency. For instance,
in \cite{928-CG-stochastIcOptimization}, the two-stage edge computing
problem with uncertainty demand is handled by CG to acquire the solution
with robustness.
\item To accelerate BP for a less elegant solution in practical applications,
branching the search tree in a greedy fashion is available, for example,
the beam search \cite{427-beamSearch} to reduce the search tree.
Besides, during the CG procedure of BP, UB and LB can be efficiently
updated after the approximation of RMP, which further expedites cutting
down the search tree.
\end{itemize}

\section{Conclusions}

This work has reviewed the basics and details of CG applied in optimization
decision problems, based on the problem structure, problem reformulation,
and Dantzig-decomposition. In order to guarantee the optimality when
applying CG to the integer programming problems, the BP mechanism
has been elaborated by building the branch tree combined with CG.
The existing applications of CG and BP in communication networks have
been summarized and analyzed, including space-air-ground networks,
D2D networks, MEC networks, body area networks, etc. It can be observed
that CG is a powerful tool to handle the optimization problems in
various communication networks to make efficient decision\textcolor{black}{.
In the applications of different communication networks with various
metrics, CG and BP can be alternatively employed according to the
specific network and computational complexity demands.}

\end{document}